\documentclass[twocolumn, twocolappendix]{aastex701}

\DeclareGraphicsExtensions{.png,.jpg,.jpeg,.pdf}
\usepackage{xcolor}
\usepackage{bm}
\usepackage{amsmath}
\usepackage{comment}

\newcommand{\vbm}{\texttt{VBMicrolensing}}

\newcommand\ltsima{$\; \buildrel <\over\sim \;$}
\newcommand\simlt{\lower.5ex\hbox{\ltsima}}
\newcommand\gtsima{$\; \buildrel >\over\sim \;$}
\newcommand\simgt{\lower.5ex\hbox{\gtsima}}

\shorttitle{}
\shortauthors{Nunota}

\graphicspath{{./}{figures/}}

\begin{document}

\title{Harmonic Angular Mapping Method for Binary Microlensing Grid Search}

\author[0009-0005-3414-455X]{Kansuke Nunota}
\affiliation{
Department of Earth and Space Science,
Graduate School of Science,
The University of Osaka,
Toyonaka, Osaka 560-0043, Japan
}
\email{nunota@iral.ess.sci.osaka-u.ac.jp}

\begin{abstract}
Searching for binary-lens microlensing solutions is computationally demanding because of the complex parameter space and the large number of trial models required.
We present an algorithm that accelerates binary-lens grid searches by expanding the angular dependence of precomputed magnification maps in Fourier modes.
By contracting the observation-dependent weighted sums in Fourier space, the $\chi^2$ values for all trial source-trajectory angles can be evaluated efficiently using inverse fast Fourier transforms (FFTs), avoiding repeated calculations over individual observations.
We tested the method on 200 Roman-like synthetic binary-lens events. For 166 events (83\%), the recovered fits agreed with Levenberg--Marquardt fits initialized at the injected parameters to within $|\Delta\chi^2|\leq1$. The FFT map scan required an average of 5.05~min per event. For the same discrete search volume, the Fourier angular evaluation was estimated to be approximately 900 times faster than conventional magnification-map sampling. These results demonstrate that the method can efficiently identify useful initial conditions for binary-lens light-curve modeling.
\end{abstract}


\section{Introduction}\label{sec:intro}

Gravitational microlensing probes the properties of foreground lens systems through the transient magnification of a background source caused by their gravitational field \citep{pac86}. Because the lensing signal does not require detectable light from the lens itself, microlensing is sensitive to faint lens stars and their planetary companions and has become an important method for detecting and characterizing planetary systems \citep{mao91,gau12}. When the lens contains a stellar or planetary companion, characteristic deviations from a single-lens light curve provide information on the properties of the lens system \citep{mao91}.

Extracting this information requires fitting the observed light curve with a binary-lens model. In many events, particularly when the binary perturbation is localized, useful initial estimates of the epoch of closest approach $t_0$, impact parameter $u_0$, and Einstein timescale $t_{\rm E}$ can be obtained from an initial single-lens fit. In contrast, suitable initial values of the mass ratio $q$, projected separation $s$, and trajectory angle $\alpha$ are generally more difficult to determine. The parameters $s$ and $q$ determine the caustic geometry, while $\alpha$ determines the orientation of the source trajectory relative to the caustic structure \citep{dom99}. Changes in $(s,q,\alpha)$ can therefore produce qualitatively different caustic crossings and cusp approaches, causing the model light curve to vary highly nonlinearly across this parameter space. The resulting $\chi^2$ surface can contain multiple separated local minima, motivating a global search over $(s,q,\alpha)$.

Several strategies have been developed to explore the complex binary-lens parameter space. One approach is to exploit characteristic light-curve morphologies to select promising initial conditions \citep{mao95,lie15}. RTModel\footnote{\url{https://github.com/valboz/RTModel}} \citep{boz24} follows this approach by using a template library of caustic-crossing and caustic-approach light curves to generate initial conditions for binary-lens fits.
An alternative is to use an ``initial-condition grid search'', in which a grid in parameters such as $(s,q,\alpha)$ is used to generate starting points for subsequent $\chi^2$ minimization, as implemented in eesunhong\footnote{\url{https://github.com/golmschenk/eesunhong}}. This approach was introduced by \citet{ben10_irs}.

In either approach, repeated calculation of binary-lens magnifications with finite-source effects \citep{wit94} can become a major computational bottleneck. For initial-condition grid searches, this cost can be reduced by precomputing magnification maps. Once the mass ratio $q$, projected separation $s$, and source radius $\rho$ are fixed, the finite-source magnification map in the source plane is also fixed and can therefore be reused for different source trajectories. MORSE \citep{mcd16} exploited this property by constructing magnification maps over a grid of binary-lens configurations and searching large numbers of source trajectories using GPUs. More recently, AdaMGrid\footnote{\url{https://github.com/zhangjiyuan22/AdaMGrid}} \citep{zha26} introduced adaptively refined magnification maps precomputed over a broad grid of $(s,q,\rho)$ and reusable for different events.

Although such magnification-map approaches avoid repeated finite-source calculations and substantially accelerate such grid searches, a conventional map-based search must still sample the precomputed map along each trial trajectory at all observation epochs and accumulate the corresponding fit statistics over the full data set. This observation-by-observation calculation is repeated for every combination of magnification map and trajectory parameters, including each trial value of the trajectory angle $\alpha$. Consequently, even with precomputed finite-source magnification maps, finding binary-lens solutions remains computationally demanding.

In this work, we develop a Fourier-based search method to accelerate this remaining step of map-based binary-lens searches. We express the angular dependence of the magnification map in Fourier modes and reorganize the calculation so that the information associated with the observation epochs, photometric weights, and source trajectory can be accumulated before scanning over the trajectory angle and reused across different precomputed maps. The dependence on the trajectory angle $\alpha$ can then be evaluated simultaneously on a uniform grid using an inverse fast Fourier transform (FFT), avoiding repeated observation-by-observation calculations for each trial angle.

The remainder of this paper is organized as follows.
Section~\ref{sec:method} describes the proposed method and its numerical implementation.
Section~\ref{sec:tests} presents tests using synthetic microlensing events.
Section~\ref{sec:disc} discusses the failure modes and computational performance.
Finally, Section~\ref{sec:summary} summarizes our findings and outlines future directions.

\section{Method}\label{sec:method}
\subsection{Binary-lens light-curve model}

As discussed in Section~\ref{sec:intro}, the finite-source magnification map in the source plane is fully specified by the mass ratio $q$, projected separation $s$, and source radius $\rho$. We denote by $(x,y)$ the source-plane coordinates expressed in units of the angular Einstein radius $\theta_{\rm E}$, and write the magnification as
\begin{equation}
A(x,y;s,q,\rho).
\label{eq:magnification_map}
\end{equation}
We choose the $x$-axis to lie along the binary axis.

The observed light curve is obtained by evaluating this magnification map along the source trajectory. Assuming rectilinear and uniform relative lens--source motion, the trajectory is specified by the epoch of closest approach $t_0$, the minimum impact parameter $u_0$, the Einstein timescale $t_{\rm E}$, and the trajectory angle $\alpha$.

Following the trajectory convention adopted by \vbm{}~\citep{boz18}, we define
\begin{equation}
\tau(t) \equiv \frac{t-t_0}{t_{\rm E}},
\label{eq:tau}
\end{equation}
and write the source position as\footnote{See the \vbm\ documentation for the binary-lens light-curve convention: \url{https://github.com/valboz/VBMicrolensing/blob/main/docs/python/LightCurves.md}.}
\begin{align}
x(t) &= u_0\sin\alpha-\tau(t)\cos\alpha,
\label{eq:source_x}\\
y(t) &= -u_0\cos\alpha-\tau(t)\sin\alpha.
\label{eq:source_y}
\end{align}
The magnification along the source trajectory is then given by
\begin{equation}
A(t) \equiv A[x(t),y(t);s,q,\rho].
\end{equation}

The observed flux for each photometric data set $j$ is modeled as
\begin{equation}
F_j(t)
=
F_{{\rm s},j} A(t)
+
F_{{\rm b},j},
\label{eq:flux_model}
\end{equation}
where $F_{{\rm s},j}$ and $F_{{\rm b},j}$ are the source and blend fluxes, respectively.
For observed fluxes $F_{i,j}$ with uncertainties $\sigma_{i,j}$, the goodness of fit is evaluated by
\begin{equation}
\chi^2
=
\sum_j \sum_i
\frac{
\left[
F_{i,j}
-
F_{{\rm s},j}A(t_{i,j})
-
F_{{\rm b},j}
\right]^2
}{
\sigma_{i,j}^2
},
\label{eq:chi2}
\end{equation}
where $i$ indexes the observations in each photometric data set. 

\subsection{Fourier-space evaluation of trajectory angles}
\label{sec:fourier_evaluation}

For a fixed $(s,q,\rho)$, changing the trajectory angle $\alpha$ corresponds to rotating the source trajectory across the same magnification map. We therefore express the magnification map in polar coordinates and expand its angular dependence in a Fourier series.
We write the source-plane coordinates as
\begin{equation}
x=r\cos\phi,
\qquad
y=r\sin\phi,
\label{eq:polar_coordinates}
\end{equation}
and define
\begin{equation}
A(r,\phi)
\equiv
A(r\cos\phi,r\sin\phi;s,q,\rho).
\label{eq:polar_magnification}
\end{equation}
The Fourier coefficients at each radius $r$ are then defined by
\begin{equation}
c_m(r)
=
\frac{1}{2\pi}
\int_0^{2\pi}
A(r,\phi)e^{-im\phi}\,d\phi.
\label{eq:fourier_coefficients}
\end{equation}
Since $A(r,\phi)$ is real, the negative Fourier modes are given by the complex conjugates of the corresponding positive modes, and only the non-negative modes need to be retained.

In practice, we truncate the Fourier expansion at a finite maximum harmonic order $M$ and write
\begin{equation}
A(r,\phi)
\simeq
A_M(r,\phi)
\equiv
c_0(r)
+
2\,{\rm Re}
\left[
\sum_{m=1}^{M}
c_m(r)e^{im\phi}
\right].
\label{eq:harmonic_expansion}
\end{equation}
Larger values of $M$ retain progressively finer angular structure in the magnification map.

For fixed values of $(t_0,u_0,t_{\rm E})$, we consider an observation at time $t_i$ and denote $\tau(t_i)$ by $\tau_i$. For $\alpha=0$, the source position can be written in polar coordinates as
\begin{equation}
r_i
=
\sqrt{\tau_i^2+u_0^2},
\qquad
\psi_i
=
\operatorname{atan2}(-u_0,-\tau_i).
\label{eq:trajectory_polar}
\end{equation}
A trajectory angle $\alpha$ simply shifts the source azimuth from $\psi_i$ to $\psi_i+\alpha$. Substituting this into Equation~(\ref{eq:harmonic_expansion}), the order-$M$ magnification at $t_i$ becomes
\begin{equation}
A_{i,M}(\alpha)
\equiv
c_0(r_i)
+
2\,{\rm Re}
\left[
\sum_{m=1}^{M}
c_m(r_i)e^{im\psi_i}e^{im\alpha}
\right].
\label{eq:trajectory_harmonic}
\end{equation}
For later convenience, we rewrite Equation~(\ref{eq:trajectory_harmonic}) in complex form by defining
\begin{equation}
\begin{aligned}
a_{i,0} &\equiv c_0(r_i),\\
a_{i,m} &\equiv c_m(r_i)e^{im\psi_i}
\quad (m>0),\\
a_{i,-m} &\equiv a_{i,m}^{*}.
\end{aligned}
\label{eq:trajectory_coefficients}
\end{equation}
Then,
\begin{equation}
A_{i,M}(\alpha)
=
\sum_{m=-M}^{M}
a_{i,m}e^{im\alpha}.
\label{eq:trajectory_harmonic_complex}
\end{equation}
Thus, for fixed $(s,q,\rho,t_0,u_0,t_{\rm E})$, the magnification is a Fourier series in the trajectory angle $\alpha$.

To exploit this representation in the evaluation of Equation~(\ref{eq:chi2}), we analytically optimize $F_{{\rm s},j}$ and $F_{{\rm b},j}$, which enter the flux model linearly. For each photometric data set $j$, all model dependence of the profiled $\chi^2$ is then contained in the following three weighted moments:
\begin{equation}
S_{A,j}(\alpha)
\equiv
\sum_i w_{i,j}A_{i,M}(\alpha),
\label{eq:SA}
\end{equation}
\begin{equation}
S_{AF,j}(\alpha)
\equiv
\sum_i w_{i,j}F_{i,j}A_{i,M}(\alpha),
\label{eq:SAF}
\end{equation}
and
\begin{equation}
S_{AA,j}(\alpha)
\equiv
\sum_i w_{i,j}A_{i,M}^2(\alpha),
\label{eq:SAA}
\end{equation}
where $w_{i,j}\equiv\sigma_{i,j}^{-2}$.

Substituting Equation~(\ref{eq:trajectory_harmonic_complex}) into Equations~(\ref{eq:SA}) and (\ref{eq:SAF}) and exchanging the sums over observations and Fourier modes gives
\begin{equation}
S_{A,j}(\alpha)
=
\sum_{m=-M}^{M}
C^{(A)}_{m,j}e^{im\alpha},
\quad
C^{(A)}_{m,j}
\equiv
\sum_i w_{i,j}a_{i,m},
\label{eq:SA_fourier}
\end{equation}
and
\begin{equation}
S_{AF,j}(\alpha)
=
\sum_{m=-M}^{M}
C^{(AF)}_{m,j}e^{im\alpha},
\quad
C^{(AF)}_{m,j}
\equiv
\sum_i w_{i,j}F_{i,j}a_{i,m}.
\label{eq:SAF_fourier}
\end{equation}
Unlike the two linear moments, the quadratic moment involves products of Fourier modes. Since the product of two Fourier series is itself a Fourier series, $A_{i,M}^2(\alpha)$ contains modes up to order $2M$. Substituting Equation~(\ref{eq:trajectory_harmonic_complex}) into Equation~(\ref{eq:SAA}) gives
\begin{equation}
S_{AA,j}(\alpha)
=
\sum_{\ell=-2M}^{2M}
C^{(AA)}_{\ell,j}e^{i\ell\alpha},
\label{eq:SAA_fourier}
\end{equation}
where
\begin{equation}
C^{(AA)}_{\ell,j}
\equiv
\sum_i w_{i,j}
\sum_{m=-M}^{M}
a_{i,m}a_{i,\ell-m},
\label{eq:SAA_coefficients}
\end{equation}
with $a_{i,m}=0$ for $|m|>M$. The inner sum represents the convolution of the Fourier coefficients.

Equations~(\ref{eq:SA_fourier}), (\ref{eq:SAF_fourier}), and (\ref{eq:SAA_fourier}) show that the three weighted moments required for the profiled $\chi^2$ are Fourier series in the trajectory angle $\alpha$. Their Fourier coefficients, $C^{(A)}_{m,j}$, $C^{(AF)}_{m,j}$, and $C^{(AA)}_{\ell,j}$, are independent of $\alpha$ and therefore need to be constructed only once for fixed $(s,q,\rho,t_0,u_0,t_{\rm E})$.

We evaluate the trajectory angle on a uniform grid,
\begin{equation}
\alpha_k
=
\frac{2\pi k}{N_\alpha},
\qquad
k=0,\ldots,N_\alpha-1,
\label{eq:alpha_grid}
\end{equation}
where $N_\alpha$ is the number of trial trajectory angles. For example, Equation~(\ref{eq:SA_fourier}) becomes
\begin{equation}
S_{A,j}(\alpha_k)
=
\sum_{m=-M}^{M}
C^{(A)}_{m,j}
\exp\left(
\frac{2\pi i m k}{N_\alpha}
\right),
\label{eq:SA_idft}
\end{equation}
which is an inverse discrete Fourier transform. The values of $S_{A,j}$ over the entire angular grid can therefore be evaluated simultaneously using an inverse fast Fourier transform (FFT; \citealt{fft}). The same procedure is applied to $S_{AF,j}$ and $S_{AA,j}$ and the resulting moments are then used to evaluate the analytically profiled $\chi^2$ at each $\alpha_k$.

An important feature of this formulation is that the summation over observations is performed before the trajectory angles are evaluated. The observation index $i$ has already been contracted into the Fourier coefficients $C^{(A)}_{m,j}$, $C^{(AF)}_{m,j}$, and $C^{(AA)}_{\ell,j}$. Consequently, changing $\alpha$ does not require recomputing the magnification and weighted sums for every observation. Instead, the observation-dependent information is first compressed into the Fourier coefficients, after which the dependence on $\alpha$ is recovered collectively by the inverse FFT.

\subsection{Numerical implementation}
\label{sec:implementation}

To numerically implement the method described above, we precompute the Fourier coefficients on a discrete grid in $(s,q,\rho)$. In the implementation, we apply the Fourier expansion to the excess magnification, $\Delta A\equiv A-1$, rather than to $A$ itself. Since the subtracted constant contributes only to the zeroth Fourier mode, this does not alter the angular structure of the expansion or the treatment of the trajectory angle $\alpha$. We adopt a parameter grid based on that used by AdaMGrid \citep{zha26}, containing $33{,}993$ combinations of $(s,q,\rho)$. For each grid point, the Fourier coefficients are computed and stored for $m=0,\ldots,512$. A lower harmonic order $M<512$ can then be used simply by truncating the stored coefficients, without recomputing the finite-source magnification.

To compute $c_m(r)$, we discretize both the radial coordinate $r$ and the azimuthal coordinate $\phi$ in Equation~(\ref{eq:fourier_coefficients}). We use 256 radial nodes, whose locations are adaptively chosen so that more nodes are placed where the Fourier spectrum varies rapidly with radius. At each radius, the number of azimuthal grid points is initially set to $N_\phi=512$ and is increased, according to the convergence of the Fourier coefficients, up to a maximum of $N_\phi=8192$. Additional angular samples are introduced locally near caustics when necessary.

For non-uniform angular sampling, the magnification between adjacent samples is approximated by a piecewise-linear function, and its contribution to the integral in Equation~(\ref{eq:fourier_coefficients}) is evaluated analytically over each interval. Since the coefficients $c_m(r)$ are stored only at the discrete radial nodes, their values at the radial source positions $r_i$ corresponding to the observations are obtained by cubic Lagrange interpolation through four neighboring radial nodes.

The remaining trajectory parameters $(t_0,u_0,t_{\rm E})$ determine the radius $r_i$ and phase $\psi_i$ of each observation through Equation~(\ref{eq:trajectory_polar}). For each trial $(t_0,u_0,t_{\rm E})$, we therefore precompute the quantities associated with the radial interpolation, phase factors, and photometric weights. These quantities are independent of $(s,q,\rho)$ and can thus be reused for all precomputed maps.

The Fourier coefficients $C^{(A)}_{m,j}$, $C^{(AF)}_{m,j}$, and $C^{(AA)}_{\ell,j}$ in Equations~(\ref{eq:SA_fourier}), (\ref{eq:SAF_fourier}), and (\ref{eq:SAA_fourier}) are then obtained by combining the map coefficients $c_m(r)$ with these trajectory-dependent quantities and summing over the radial nodes. This calculation is performed in batches over multiple $(s,q,\rho)$ maps and trial $(t_0,u_0,t_{\rm E})$ trajectories, yielding the Fourier coefficients for each map--trajectory combination. Inverse FFTs along the Fourier-mode direction are then used to evaluate $S_{A,j}$, $S_{AF,j}$, and $S_{AA,j}$ over the grid of trajectory angles $\alpha$, from which the profiled $\chi^2$ is calculated. These batched operations and inverse FFTs are implemented with JAX \citep{jax18} and compiled using just-in-time (JIT) compilation.
The algorithm is publicly available as the Python package
\texttt{hamagri}\footnote{\url{https://github.com/NunotaKansuke/hamagri}}.

\section{Tests}
\label{sec:tests}

We evaluate the performance of the proposed method using synthetic binary-lens light curves. The tests are designed to examine two aspects of the method: whether the grid search can recover the relevant binary-lens solutions from realistically sampled and noisy light curves, and how much our method reduces the computational cost of the search.

\subsection{Test setup}
\label{sec:test_setup}

We generated synthetic light curves using the GULLS microlensing survey simulator~\citep{pen19}. We adopted the Roman WFI F146 configuration for field 0 of the Roman Microlensing Data Challenge 2026\footnote{\url{https://rges-pit.org/data-challenge/}}. The simulations span 859 days over three observing seasons and contain 23,104 epochs for each event. Each exposure has an integration time of 66.9~s, with a nominal cadence of 728.508~s, while the actual observing times include the seasonal gaps.

The source, lens, and surrounding stellar populations were generated using \texttt{genstars}~\citep{kos22}. The physical properties of each source--lens pair determined $\theta_{\rm E}$, $t_{\rm E}$, and $\rho$. The binary-lens parameters were chosen to test the grid search over a broad region of parameter space rather than to reproduce an astrophysical binary population. 
Therefore, the recovery fractions reported below characterize the performance on this test sample and should not be interpreted as recovery efficiencies for an astrophysical population.
We generated 200 synthetic events spanning $10^{-6}\lesssim q\lesssim0.957$ and $0.304\lesssim s\lesssim2.992$, including close, resonant, and wide configurations. The parameters $u_0$, $\alpha$, and $t_0$ were varied to include a range of source trajectories, with some events generated using a heuristic designed to increase the occurrence of caustic-sensitive geometries. The events were selected to exhibit a detectable binary-lens signal, with $\Delta\chi^2 \geq 100$ relative to a single-lens model. We assumed static binary lenses with uniform finite sources and neglected microlens parallax and lens orbital motion.

For each event, we used 16 trial combinations of $(t_0,u_0,t_{\rm E})$ centered on the injected trajectory parameters. The characteristic widths of this grid were
\begin{align}
&\Delta t_0=\max(2~{\rm day},\,0.1t_{\rm E}),\notag\\
&\Delta u_0=\max(0.03,\,0.1|u_0|),\notag\\
&\Delta\log t_{\rm E}=0.15\notag.
\end{align}

For each trajectory, we searched $33{,}993$ precomputed $(s,q,\rho)$ maps. The Fourier expansion was truncated at $M=128$, and the trajectory angle $\alpha$ was evaluated at 2048 uniformly spaced points over $0\leq\alpha<2\pi$. The search for each event therefore covers approximately
\[
16\times2048\times33{,}993
\simeq 1.11\times10^9
\]
grid points. 

From the FFT search, we selected 24 promising local minima as initial candidates. Each candidate was refined using Nelder--Mead optimization with direct \texttt{VBMicrolensing} evaluations, allowing up to 64 function evaluations. The 12 candidates with the lowest $\chi^2$ were then further refined using Levenberg--Marquardt (LM) optimization with a maximum of 100 function evaluations. The solution with the lowest final $\chi^2$ was adopted.

These calculations were performed on a dual-socket system equipped with Intel Xeon Gold 6530 processors using three CPU workers.

\subsection{Results}
\label{sec:test_results}

\begin{figure*}
\centering
\includegraphics[width=0.9\textwidth]{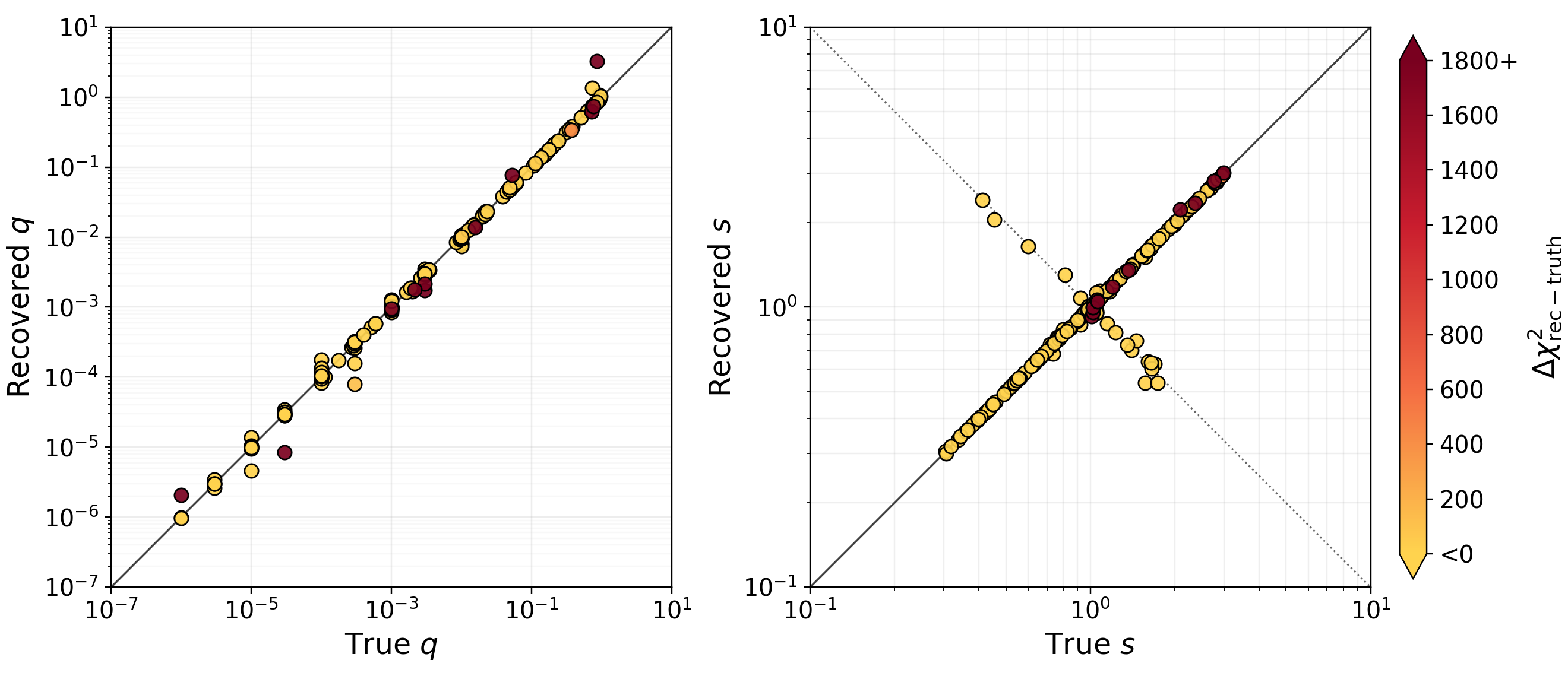}
\caption{Comparison of the injected and recovered binary-lens parameters. The left and right panels show the mass ratio $q$ and projected separation $s$, respectively. The solid lines indicate the one-to-one relations, while the dotted line in the right panel shows $s_{\rm rec}=1/s_{\rm true}$. Points are colored by the difference in $\chi^2$ between the recovered and optimized truth fits, $\Delta\chi^2_{\rm rec-truth}$.}
\label{fig:qs_recovery}
\end{figure*}

We compared the injected parameters with the recovered solutions for the 200 synthetic events.
For a diagnostic comparison, we also optimized all seven binary-lens parameters using LM, starting from the injected values. We denote the resulting fit statistic by $\chi^2_{\rm truth}$ and define $\Delta\chi^2_{\rm rec-truth}=\chi^2_{\rm recovered}-\chi^2_{\rm truth}$.
Figure~\ref{fig:qs_recovery} shows the recovered mass ratio $q$ and separation $s$ as a function of their injected values, with points colored by $\Delta\chi^2_{\rm rec-truth}$. 
Overall, the recovered parameters closely follow the injected values or, in $s$, the corresponding close--wide degenerate branch, $s_{\rm rec}\simeq1/s_{\rm true}$.

\begin{figure}
\centering
\includegraphics[width=1\columnwidth]{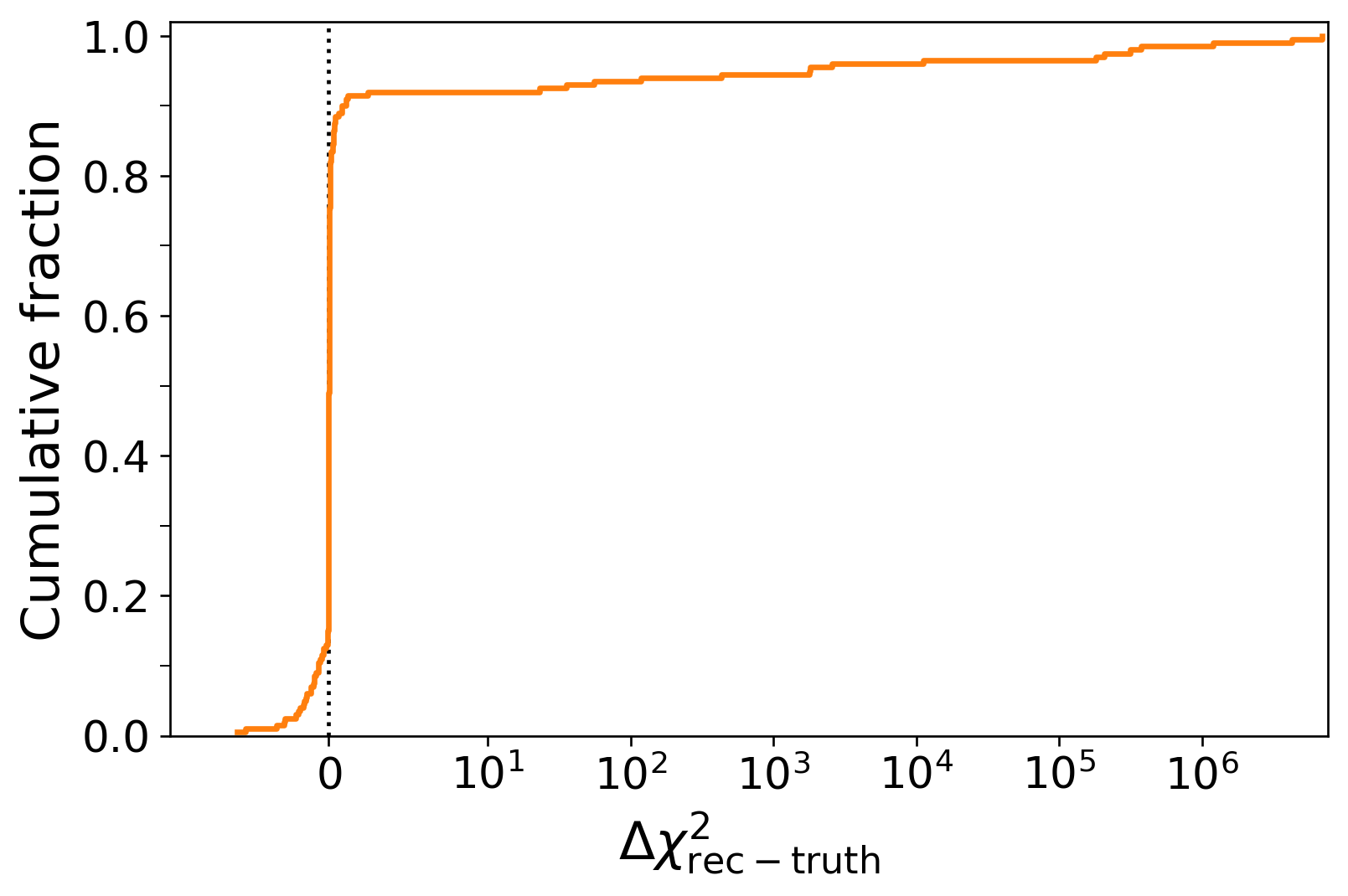}
\caption{Cumulative distribution of $\Delta\chi^2_{\rm rec-truth}$ for the 200 synthetic events. The vertical dotted line marks $\Delta\chi^2_{\rm rec-truth}=0$.}
\label{fig:dchi2_cumulative}
\end{figure}

Figure~\ref{fig:dchi2_cumulative} shows the cumulative distribution of $\Delta\chi^2_{\rm rec-truth}$. The recovered and optimized truth fits agree to within $|\Delta\chi^2_{\rm rec-truth}|\leq1$ for 166 of 200 events (83\%) and within 10 for 184 events (92\%). Although these fits are not guaranteed to represent the global minimum, the close agreement indicates that the Fourier search can provide initial conditions leading to fits comparable in quality to those obtained from the true parameters. Representative cases with larger differences are discussed in Section~\ref{sec:failure_modes}.

For the 200 synthetic events, the FFT map scan with $M=128$, $N_\alpha=2048$, and 16 trial trajectories required an average wall-clock time of 5.05~min per event (median 4.78~min; range 1.76--9.11~min).

\begin{figure*}
\centering
\includegraphics[width=0.9\textwidth]{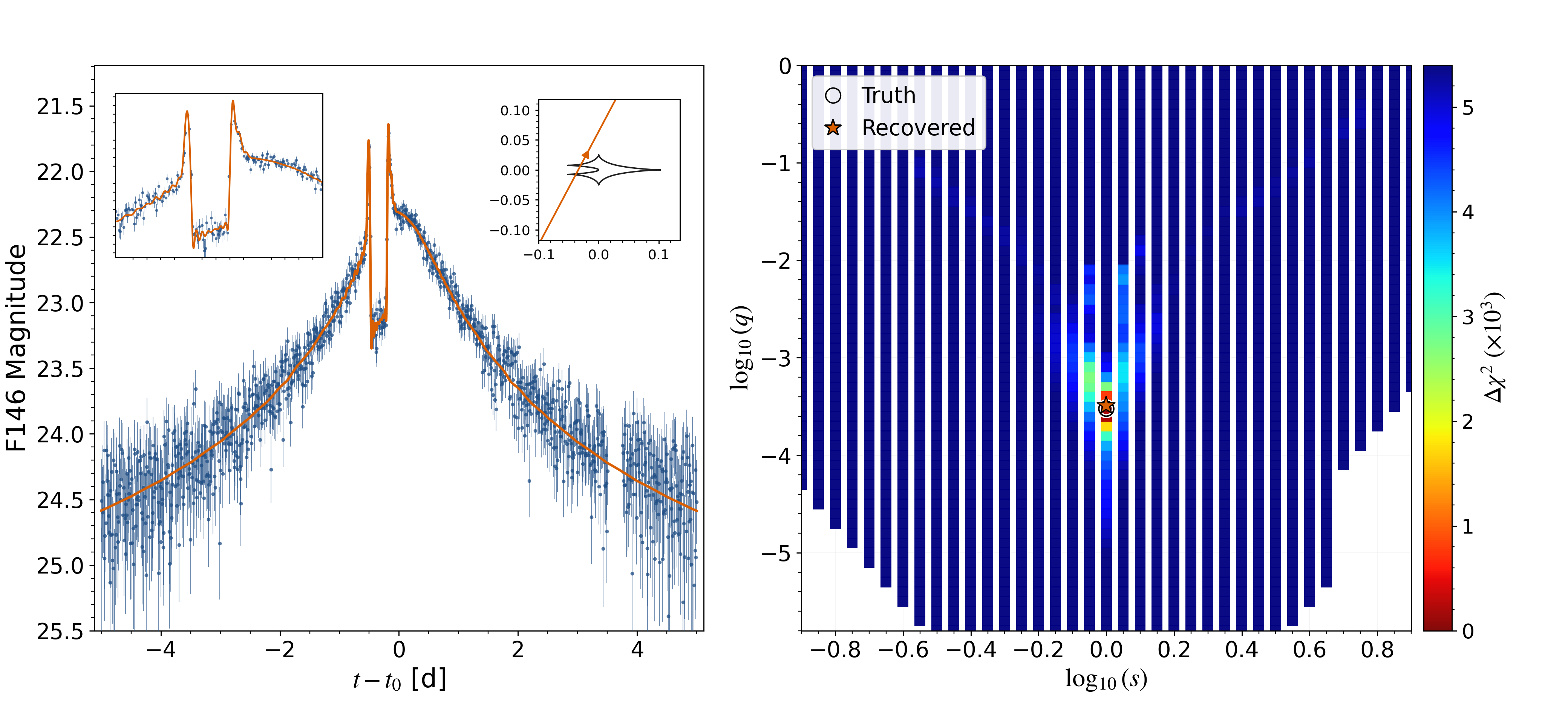}
\caption{Representative search result for a synthetic event with an isolated solution. The left panel shows the simulated light curve together with the recovered binary-lens model; the enlarged inset highlights the caustic-crossing feature, and the second inset shows the corresponding caustic geometry and source trajectory. The right panel shows the $\Delta\chi^2$ map in the $(s,q)$ plane relative to the best-fitting grid solution, with the injected parameters marked by the open circle.}
\label{fig:example_clean}
\end{figure*}

\begin{figure*}
\centering
\includegraphics[width=0.9\textwidth]{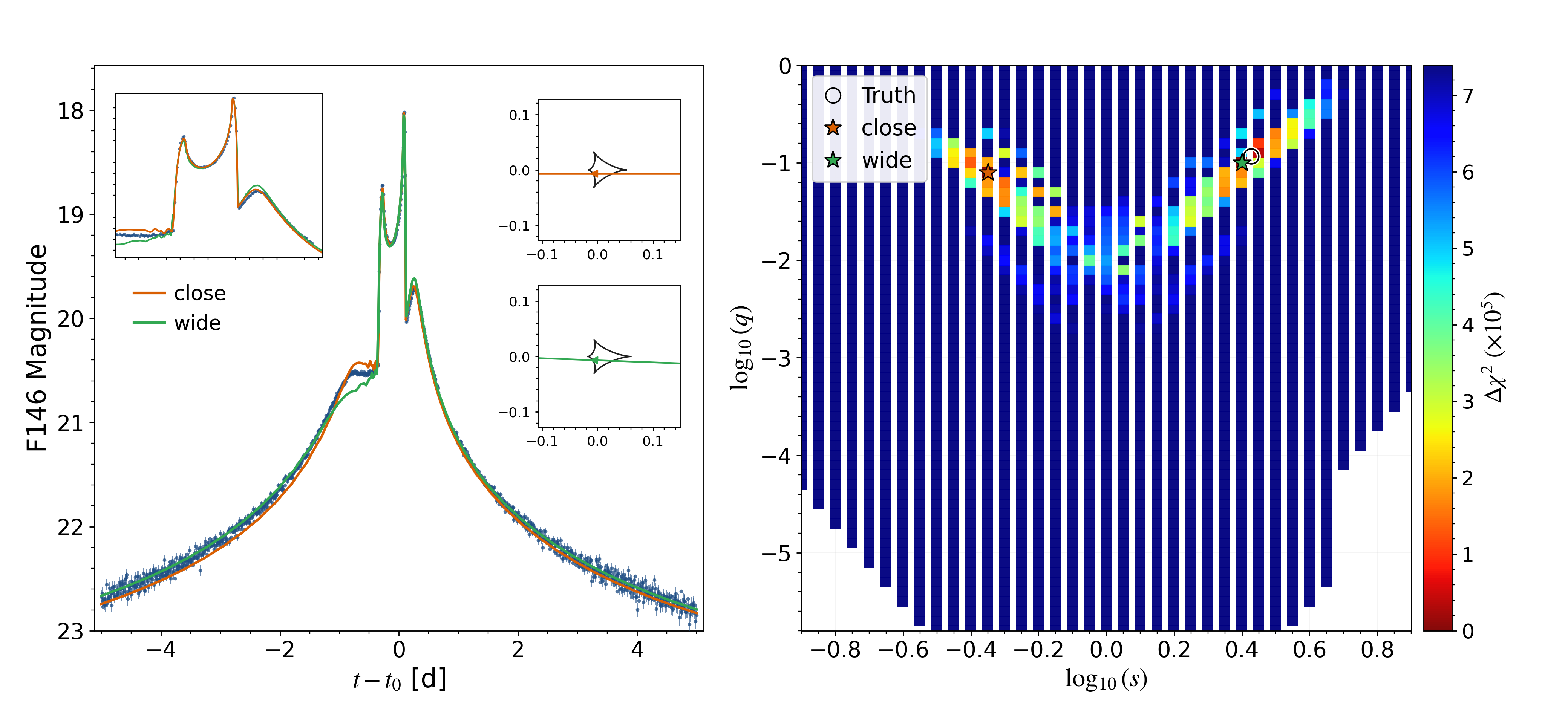}
\caption{Representative search result for a synthetic event with close--wide degenerate solutions, shown in the same format as Figure~\ref{fig:example_clean}. The recovered close and wide solutions are shown in orange and green, respectively, and correspond to distinct low-$\chi^2$ regions in the $(s,q)$ plane.}
\label{fig:example_closewide}
\end{figure*}

Figure~\ref{fig:example_clean} shows a representative event for which the search recovers an isolated minimum close to the injected $(s,q)$ parameters. Figure~\ref{fig:example_closewide} shows a representative event with both close and wide solutions. In this case, the search identifies multiple low-$\chi^2$ regions in the $(s,q)$ plane, and both the close and wide models provide comparably good descriptions of the simulated light curve. The small oscillatory features visible in the FFT light-curve models arise from the finite-order Fourier representation and are numerical approximation errors rather than physical light-curve structure.

\section{Discussion}
\label{sec:disc}

\subsection{Failure modes}
\label{sec:failure_modes}

\begin{figure*}
\centering
\includegraphics[width=0.95\textwidth]{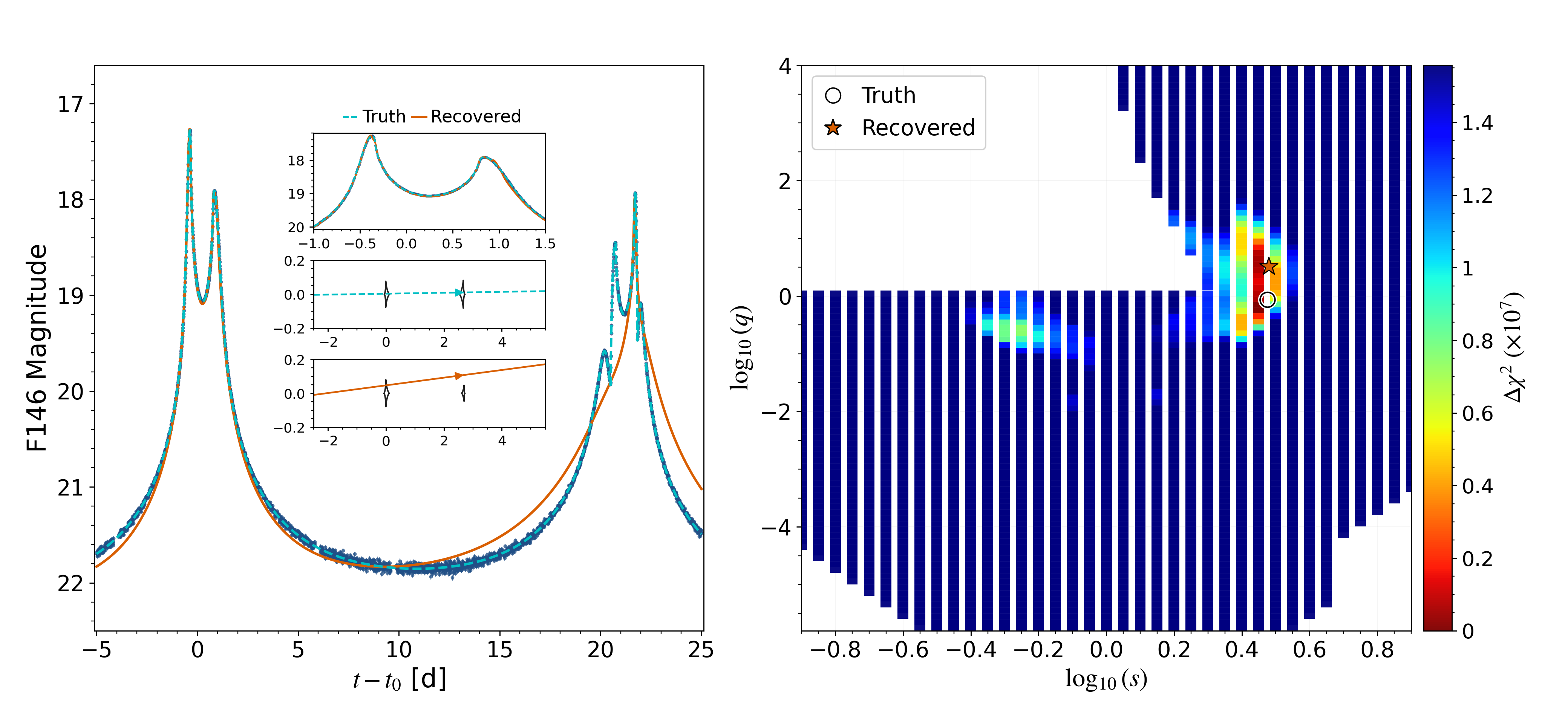}
\caption{Representative failure case in the wide-separation regime, shown in the same format as Figure~\ref{fig:example_clean}. The injected truth is shown by the cyan dashed curve, and the recovered solution by the orange curve.}
\label{fig:example_failure}
\end{figure*}

\begin{figure*}
\centering
\includegraphics[width=0.95\textwidth]{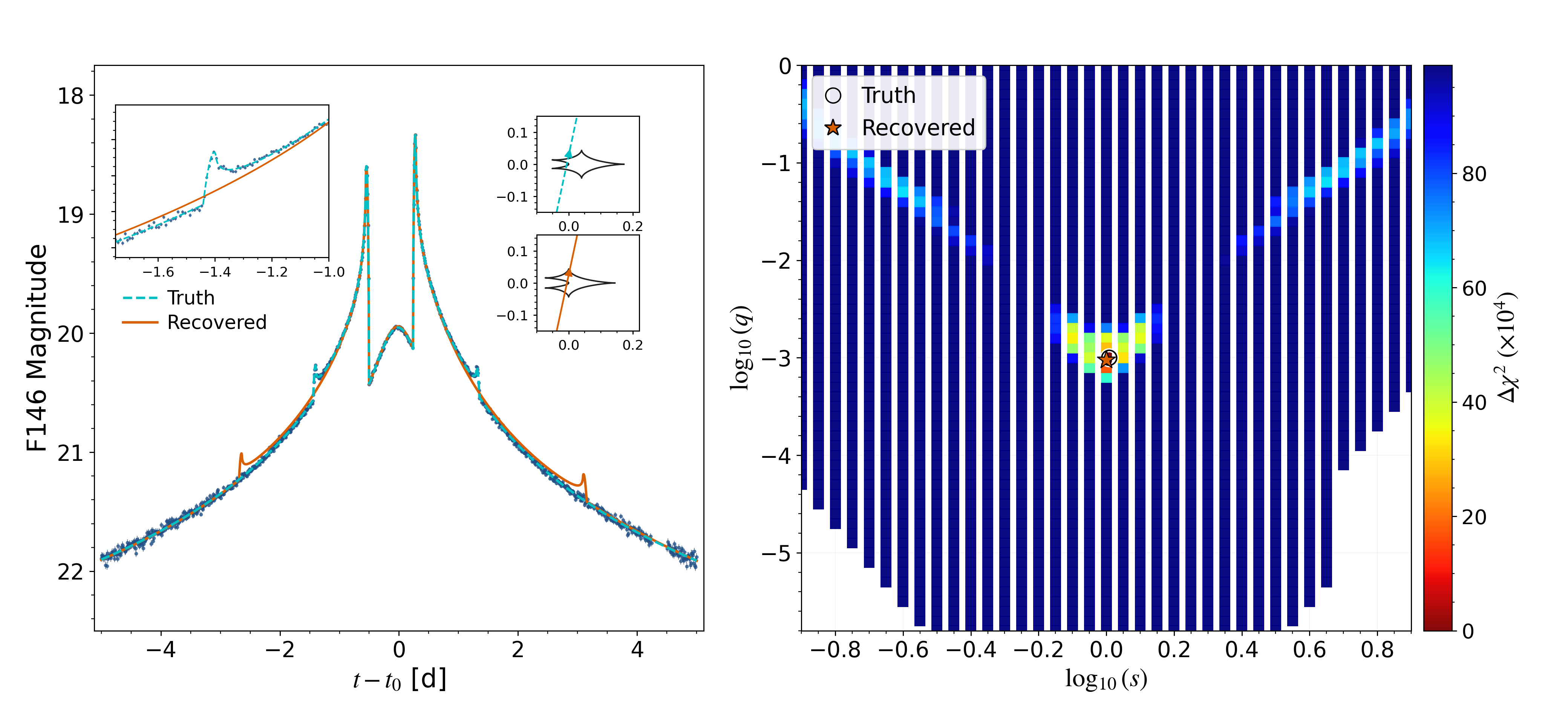}
\caption{Representative failure case in the resonant regime, shown in the same format as Figure~\ref{fig:example_failure}.}
\label{fig:resonant_failure}
\end{figure*}

Two characteristic difficult configurations in our tests are wide-binary events in which the source probes spatially separated caustic regions and events involving a resonant caustic near $s\simeq1$. These are illustrative failure modes rather than an exhaustive classification of the 200-event sample.

Figure~\ref{fig:example_failure} shows an example of the first type.
For wide binaries, the two caustic regions become increasingly separated and smaller as $s$ increases \citep{dom99}.
In our maps, the coordinate system is centered on one of the caustics, so the other appears as a small structure far from the origin and is confined to a narrow range in azimuth.
Such localized angular structure is difficult to reproduce accurately with a finite Fourier expansion, introducing errors in the FFT $\chi^2$.
As a result, the Fourier search can select an initial solution that reproduces the caustic feature near the origin but not the distant one.
Starting from such a solution, the subsequent local optimization may remain trapped in a local minimum that fits one feature well without simultaneously recovering the other.

Figure~\ref{fig:resonant_failure} shows an example of the second type, which occurs mainly for events involving resonant caustics near $s\simeq1$.
In this regime, the caustic structure can change rapidly with $s$ \citep{chu05}, making the subsequent local optimization sensitive to the initial value of $s$.
For the event shown, changing only $s$ from its injected value of 1.02 to the neighboring grid value of 1.00 caused the local LM optimization to converge to a different solution with $\Delta\chi^2=11{,}612.35$.
Thus, the combination of this strong sensitivity to the initial $s$ and the finite spacing of the $s$ grid can cause the global solution to be missed.

The relative frequency of these failure modes in our deliberately selected sample should not be interpreted as their occurrence rates in an astrophysical population. For planetary events, the resonant failure mode is likely to be more relevant, whereas wide-separation events in which the source probes both caustic regions are rare, with examples including OGLE-2008-BLG-092 and OGLE-2016-BLG-0263 \citep{pol14,han17}.

In both failure modes, the recovered solutions generally remain close to the relevant solution region rather than being displaced to unrelated parts of parameter space. 
Thus, although a more robust local refinement strategy may be needed in these difficult cases, the Fourier search still serves its primary purpose as a rapid search for useful initial conditions.

\subsection{Computational performance}
\label{sec:performance}

We evaluated the runtime performance of the Fourier search by measuring its dependence on the number of trajectory angles $N_\alpha$, the harmonic order $M$, the number of trial $(t_0,u_0,t_{\rm E})$ combinations, and the number of CPU workers.

\paragraph{Trajectory-angle resolution.}
To examine the dependence on the trajectory-angle resolution, we repeated the search described in Section~\ref{sec:test_setup} for the event shown in Figure~\ref{fig:example_clean}, using four CPU workers and varying the number of trial angles over $N_\alpha=1024,2048,$ and $4096$.
The corresponding runtimes were 155.6, 173.2, and 176.8~s, respectively. Thus, increasing $N_\alpha$ by a factor of four increased the runtime by only about 14\%, while doubling it from 2048 to 4096 increased the runtime by only about 2\%. The recovered solution was unchanged over this range of $N_\alpha$.

This weak dependence arises because the observation-dependent weighted sums are contracted into Fourier coefficients before the trajectory angles are evaluated. Increasing $N_\alpha$ therefore does not require repeating the observation-by-observation map evaluation or Fourier contraction. The $N_\alpha$-dependent work is dominated by the inverse FFTs and the subsequent evaluation of $\chi^2$ over the angular grid, with approximate costs of $\mathcal{O}(N_\alpha\log N_\alpha)$ and $\mathcal{O}(N_\alpha)$, respectively. Over the range tested here, these operations constitute only a small fraction of the total runtime, allowing a dense trajectory-angle grid to be used at relatively little additional cost.

\paragraph{Fourier resolution.}
To examine the dependence on the Fourier resolution, we repeated the same search with $N_\alpha=2048$ while varying the harmonic order over $M=32,64,128,$ and $256$. The corresponding runtimes were 56.1, 87.7, 173.2, and 318.8~s, respectively, showing a clear increase in computational cost with increasing $M$. The same solution was recovered for all four values of $M$ in this event.

However, the harmonic order required for an accurate search is expected to depend on the event morphology and caustic structure. Therefore, this single-event test does not imply that lower values of $M$ are generally sufficient. In particular, more localized angular structures, such as the wide-binary caustics discussed in Section~\ref{sec:failure_modes}, may require a higher Fourier resolution.

\paragraph{Number of $(t_0,u_0,t_{\rm E})$ trials.}
For real events, the initial values of $(t_0,u_0,t_{\rm E})$ cannot always be estimated accurately. In particular, when a planetary anomaly or caustic feature is strong, a simple single-lens fit can be biased by the perturbation and may not provide adequate initial estimates of $(t_0,u_0,t_{\rm E})$. In such cases, it is useful to search a broader grid in $(t_0,u_0,t_{\rm E})$.

To examine the runtime dependence on the trajectory-grid size, we repeated the same search while increasing the number of trial $(t_0,u_0,t_{\rm E})$ combinations from 16 to 32, 64, and 128. The corresponding runtimes were 173.2, 234.5, 595.1, and 1022.5~s, respectively. Increasing the number of trial combinations by a factor of eight therefore increased the runtime by a factor of about 5.9, indicating that the computational cost does not scale strictly linearly with the number of trials over the range tested.

This sublinear increase is consistent with the reuse of the same map coefficients across multiple $(t_0,u_0,t_{\rm E})$ trials and the batched evaluation of these trials which allows part of the computational cost to be shared or amortized. Thus, when the initial values of $(t_0,u_0,t_{\rm E})$ are uncertain, broadening the $(t_0,u_0,t_{\rm E})$ grid remains computationally practical.

\paragraph{Number of CPU workers.}
For larger-scale analyses, the grid search may be run using more CPU workers than the four used in the preceding benchmarks. To quantify the parallel scaling, we repeated the same search using 1, 4, 8, and 16 CPU workers. The corresponding runtimes were 602.9, 173.2, 90.9, and 66.4~s, giving speedups of 3.48, 6.63, and 9.08 relative to the single-worker run. The corresponding parallel efficiencies were approximately 87\%, 83\%, and 57\%, respectively, showing that the efficiency decreases as more workers are used.

This loss of efficiency likely reflects both fixed costs that are not fully parallelized, such as worker initialization and result aggregation, and increasing contention for shared resources such as memory bandwidth and cache when multiple workers process the same coefficient data. Thus, increasing the number of workers continues to reduce the wall-clock time, but with progressively smaller gains at higher worker counts.

\paragraph{Comparison with conventional map sampling.}
Finally, we estimated the computational gain obtained by introducing the Fourier representation relative to conventional magnification-map sampling. Using the standard map interpolation in AdaMGrid~\citep{zha26}, we measured the runtime for 10 maps and one fixed combination of $(t_0,u_0,t_{\rm E})$, and linearly extrapolated the result to the scale of our standard search. The extrapolated conventional calculation is estimated to require approximately 44~hr, compared with 173.2~s for the Fourier map scan in our four-worker benchmark, corresponding to a speedup of about $900$ over the same discrete search volume.

However, AdaMGrid performs more than a fixed-grid scan: it can continuously optimize the trajectory parameters for each map and employ MCMC for more robust local exploration~\citep{zha26}.

The estimated factor of $900$ therefore applies to the discrete map evaluation rather than the overall performance or robustness of the two search methods.

\section{Summary and Future Directions}
\label{sec:summary}

\subsection{Summary}

We have developed an algorithm for accelerating binary-lens grid searches by expanding precomputed magnification maps in Fourier modes and evaluating the trajectory-angle dependence with inverse FFTs. In this formulation, the observation-dependent weighted sums are contracted before scanning over the source-trajectory angle $\alpha$, avoiding repeated observation-by-observation map sampling for each trial angle.

We tested the method on 200 Roman-like synthetic events. For 166 events (83\%), the recovered and optimized truth fits differed by $|\Delta\chi^2_{\rm rec-truth}|\leq1$, and 184 events (92\%) were within 10. Although these diagnostic comparisons do not guarantee that a global minimum has been found, the close agreement for most events supports the method's usefulness for identifying initial conditions.

The method also provides a substantial computational advantage. Increasing $N_\alpha$ from 1024 to 4096 changed the runtime by only about 14\%, allowing a dense angular grid to be searched at relatively little additional cost. The FFT map scan required an average wall-clock time of 5.05~min per event. For the same discrete search volume, comparison with conventional magnification-map sampling indicates an estimated computational gain of approximately a factor of 900.

\subsection{Future directions}

Several aspects remain to be improved for applications to real events. In the present test, the grid in $(t_0,u_0,t_{\rm E})$ was centered on the injected values, whereas these parameters may be substantially less well constrained in real light curves, particularly when strong caustic features bias an initial single-lens fit. More robust initialization will therefore be important for a fully blind search. At the same time, our runtime tests show that increasing the number of $(t_0,u_0,t_{\rm E})$ trials is computationally feasible, suggesting that a broader grid can be used when the initial estimates are uncertain.

The framework can also be extended to include microlens parallax. 
For fixed $(t_0,u_0,t_{\rm E},\pi_{{\rm E},N},\pi_{{\rm E},E})$, the parallax perturbation modifies the source trajectory at each epoch, but the dependence on the trajectory angle $\alpha$ can still be treated with the same Fourier formalism.
The additional parallax parameters can therefore be included as extra dimensions of the trajectory grid and evaluated in batches in the same way as $(t_0,u_0,t_{\rm E})$. 
This, however, further increases the dimensionality of the search and makes the choice of initial estimates and grid placement for these parameters more important.
Lens orbital motion is less straightforward because changes in the projected separation $s$ alter the magnification map itself and cannot be incorporated directly into the present static-map framework.

There is also room to improve the resolution of the global search itself. 
The present calculations use fixed grids in $(s,q,\rho)$ and a fixed harmonic order $M$, although the required resolution is not uniform across parameter space. The failure modes discussed in Section~\ref{sec:failure_modes} suggest that wide configurations may benefit from higher Fourier resolution, while the resonant regime may require finer sampling in $s$. Adaptive refinement of these quantities could therefore improve the recovery while retaining most of the computational advantage of the initial search.

Further improvement is also possible in the subsequent local optimization. In several failed cases, the Fourier search identifies a solution close to the relevant region of parameter space, but the local optimization converges to a nearby local minimum. A more robust refinement strategy may therefore recover these cases.
More generally, the Fourier search can be used as a rapid initial-condition finder and combined with existing modeling frameworks such as RTModel \citep{boz24}, AdaMGrid \citep{zha26} or the initial-condition grid-search approach \citep{ben10_irs} for more detailed optimization.

\begin{acknowledgments}
I thank Jon Hulberg for helpful comments on the manuscript.
I used AI-assisted tools, including ChatGPT and Codex (OpenAI), to assist with software development and manuscript preparation. I take full responsibility for the scientific content and conclusions of this work.
This work is supported by the Japan Society for the Promotion of Science (JSPS)
Research Fellowship for Young Scientists (DC) and by the Next-Generation
Global Leaders Overseas Training Program of The University of Osaka.
\end{acknowledgments}

\software{
\texttt{hamagri}, 
JAX \citep{jax18},
VBMicrolensing \citep{boz18},
GULLS \citep{pen19},
genstars \citep{kos22},
AdaMGrid \citep{zha26}
}

\bibliographystyle{aasjournalv7}
\bibliography{reference}

@ARTICLE{pac86,
       author = {{Paczynski}, B.},
        title = "{Gravitational Microlensing by the Galactic Halo}",
      journal = {\apj},
         year = 1986,
        month = may,
       volume = {304},
        pages = {1-5},
          doi = {10.1086/164079},
       adsurl = {https://ui.adsabs.harvard.edu/abs/1986ApJ...304....1P}
}

@ARTICLE{mao91,
       author = {{Mao}, Shude and {Paczynski}, Bohdan},
        title = "{Gravitational Microlensing by Double Stars and Planetary Systems}",
      journal = {\apjl},
         year = 1991,
        month = jun,
       volume = {374},
        pages = {L37-L40},
          doi = {10.1086/186066},
       adsurl = {https://ui.adsabs.harvard.edu/abs/1991ApJ...374L..37M}
}

@ARTICLE{ben10_irs,
       author = {{Bennett}, David P.},
        title = "{An Efficient Method for Modeling High-magnification Planetary Microlensing Events}",
      journal = {\apj},
         year = 2010,
        month = jun,
       volume = {716},
       number = {2},
        pages = {1408-1422},
          doi = {10.1088/0004-637X/716/2/1408},
archivePrefix = {arXiv},
       eprint = {0911.2703},
 primaryClass = {astro-ph.EP},
       adsurl = {https://ui.adsabs.harvard.edu/abs/2010ApJ...716.1408B}
}

@ARTICLE{boz18,
       author = {{Bozza}, Valerio and {Bachelet}, Etienne and {Bartoli{\'c}}, Fran and {Heintz}, Tyler M. and {Hoag}, Ava R. and {Hundertmark}, Markus},
        title = "{VBBinaryLensing: a Public Package for Microlensing Light-curve Computation}",
      journal = {\mnras},
         year = 2018,
        month = oct,
       volume = {479},
       number = {4},
        pages = {5157-5167},
          doi = {10.1093/mnras/sty1791},
archivePrefix = {arXiv},
       eprint = {1805.05653},
 primaryClass = {astro-ph.IM},
       adsurl = {https://ui.adsabs.harvard.edu/abs/2018MNRAS.479.5157B}
}

@ARTICLE{wit94,
       author = {{Witt}, Hans J. and {Mao}, Shude},
        title = "{Can Lensed Stars Be Regarded as Pointlike for Microlensing by MACHOs?}",
      journal = {\apj},
         year = 1994,
        month = aug,
       volume = {430},
        pages = {505},
          doi = {10.1086/174426},
       adsurl = {https://ui.adsabs.harvard.edu/abs/1994ApJ...430..505W}
}

@ARTICLE{boz24,
       author = {{Bozza}, V.},
        title = "{RTModel: A platform for real-time modeling and massive analyses of microlensing events}",
      journal = {\aap},
         year = 2024,
        month = aug,
       volume = {688},
          eid = {A83},
        pages = {A83},
          doi = {10.1051/0004-6361/202450450},
archivePrefix = {arXiv},
       eprint = {2405.04092},
 primaryClass = {astro-ph.IM},
       adsurl = {https://ui.adsabs.harvard.edu/abs/2024A&A...688A..83B}
}

@ARTICLE{mcd16,
       author = {{McDougall}, Alistair and {Albrow}, Michael D.},
        title = "{Microlensing observations rapid search for exoplanets: MORSE code for GPUs}",
      journal = {\mnras},
         year = 2016,
        month = feb,
       volume = {456},
       number = {1},
        pages = {565-570},
          doi = {10.1093/mnras/stv2609},
archivePrefix = {arXiv},
       eprint = {1511.01157},
 primaryClass = {astro-ph.IM},
       adsurl = {https://ui.adsabs.harvard.edu/abs/2016MNRAS.456..565M}
}

@software{zha26,
  author = {Zhang, Jiyuan},
  title = {{AdaMGrid}},
  year = {2026},
  month = feb,
  url = {https://github.com/zhangjiyuan22/AdaMGrid}
}

@ARTICLE{fft,
       author = {{Cooley}, James W. and {Tukey}, John W.},
        title = "{An algorithm for the machine calculation of complex Fourier series}",
      journal = {Mathematics of Computation},
         year = 1965,
        month = jan,
       volume = {19},
       number = {90},
        pages = {297-301},
          doi = {10.1090/S0025-5718-1965-0178586-1},
       adsurl = {https://ui.adsabs.harvard.edu/abs/1965MaCom..19..297C}
}

@ARTICLE{gau12,
       author = {{Gaudi}, B. Scott},
        title = "{Microlensing Surveys for Exoplanets}",
      journal = {\araa},
         year = 2012,
        month = sep,
       volume = {50},
        pages = {411-453},
          doi = {10.1146/annurev-astro-081811-125518},
       adsurl = {https://ui.adsabs.harvard.edu/abs/2012ARA&A..50..411G}
}

@ARTICLE{dom99,
       author = {{Dominik}, M.},
        title = "{The binary gravitational lens and its extreme cases}",
      journal = {\aap},
         year = 1999,
        month = sep,
       volume = {349},
        pages = {108-125},
          doi = {10.48550/arXiv.astro-ph/9903014},
archivePrefix = {arXiv},
       eprint = {astro-ph/9903014},
 primaryClass = {astro-ph},
       adsurl = {https://ui.adsabs.harvard.edu/abs/1999A&A...349..108D}
}

@ARTICLE{mao95,
       author = {{Mao}, S. and {Di Stefano}, R.},
        title = "{Interpretation of Gravitational Microlensing by Binary Systems}",
      journal = {\apj},
         year = 1995,
        month = feb,
       volume = {440},
        pages = {22},
          doi = {10.1086/175244},
       adsurl = {https://ui.adsabs.harvard.edu/abs/1995ApJ...440...22M}
}

@ARTICLE{lie15,
       author = {{Liebig}, Christine and {D'Ago}, Giuseppe and {Bozza}, Valerio and {Dominik}, Martin},
        title = "{The complete catalogue of light curves in equal-mass binary microlensing}",
      journal = {\mnras},
         year = 2015,
        month = jun,
       volume = {450},
       number = {2},
        pages = {1565-1584},
          doi = {10.1093/mnras/stv733},
archivePrefix = {arXiv},
       eprint = {1501.02219},
 primaryClass = {astro-ph.EP},
       adsurl = {https://ui.adsabs.harvard.edu/abs/2015MNRAS.450.1565L}
}

@ARTICLE{pen19,
       author = {{Penny}, Matthew T. and {Gaudi}, B. Scott and {Kerins}, Eamonn and {Rattenbury}, Nicholas J. and {Mao}, Shude and {Robin}, Annie C. and {Calchi Novati}, Sebastiano},
        title = "{Predictions of the WFIRST Microlensing Survey. I. Bound Planet Detection Rates}",
      journal = {\apjs},
         year = 2019,
        month = mar,
       volume = {241},
       number = {1},
          eid = {3},
        pages = {3},
          doi = {10.3847/1538-4365/aafb69},
archivePrefix = {arXiv},
       eprint = {1808.02490},
 primaryClass = {astro-ph.EP},
       adsurl = {https://ui.adsabs.harvard.edu/abs/2019ApJS..241....3P}
}

@misc{kos22,
  author       = {Koshimoto, Naoki},
  title        = {genstars: A Tool for Stellar Population Synthesis toward the Galactic Bulge},
  year         = {2022},
  publisher    = {Zenodo},
  version      = {v1.0.0},
  doi          = {10.5281/zenodo.6941087},
  url          = {https://doi.org/10.5281/zenodo.6941087}
}

@ARTICLE{chu05,
       author = {{Chung}, Sun-Ju and {Han}, Cheongho and {Park}, Byeong-Gon and {Kim}, Doeon and {Kang}, Sangjun and {Ryu}, Yoon-Hyun and {Kim}, Kang Min and {Jeon}, Young-Beom and {Lee}, Dong-Wook and {Chang}, Kyongae and {Lee}, Woo-Baik and {Kang}, Yong Hee},
        title = "{Properties of Central Caustics in Planetary Microlensing}",
      journal = {\apj},
         year = 2005,
        month = sep,
       volume = {630},
       number = {1},
        pages = {535-542},
          doi = {10.1086/432048},
archivePrefix = {arXiv},
       eprint = {astro-ph/0505363},
 primaryClass = {astro-ph},
       adsurl = {https://ui.adsabs.harvard.edu/abs/2005ApJ...630..535C}
}

@ARTICLE{pol14,
       author = {{Poleski}, Rados{\l}aw and {Skowron}, Jan and {Udalski}, Andrzej and {Han}, Cheongho and {Koz{\l}owski}, Szymon and {Wyrzykowski}, {\L}ukasz and {Dong}, Subo and {Szyma{\'n}ski}, Micha{\l} K. and {Kubiak}, Marcin and {Pietrzy{\'n}ski}, Grzegorz and {Soszy{\'n}ski}, Igor and {Ulaczyk}, Krzysztof and {Pietrukowicz}, Pawe{\l} and {Gould}, Andrew},
        title = "{Triple Microlens OGLE-2008-BLG-092L: Binary Stellar System with a Circumprimary Uranus-type Planet}",
      journal = {\apj},
         year = 2014,
        month = nov,
       volume = {795},
       number = {1},
          eid = {42},
        pages = {42},
          doi = {10.1088/0004-637X/795/1/42},
archivePrefix = {arXiv},
       eprint = {1408.6223},
 primaryClass = {astro-ph.EP},
       adsurl = {https://ui.adsabs.harvard.edu/abs/2014ApJ...795...42P}
}

@ARTICLE{han17,
       author = {{Han}, C. and {Udalski}, A. and {Gould}, A. and {Bond}, I.~A. and {and} and {Albrow}, M.~D. and {Chung}, S.-J. and {Jung}, Y.~K. and {Ryu}, Y.-H. and {Shin}, I.-G. and {Yee}, J.~C. and {Zhu}, W. and {Cha}, S.-M. and {Kim}, S.-L. and {Kim}, D.-J. and {Lee}, C.-U. and {Lee}, Y. and {Park}, B.-G. and {KMTNet Collaboration} and {Skowron}, J. and {Mr{\'o}z}, P. and {Pietrukowicz}, P. and {Koz{\l}owski}, S. and {Poleski}, R. and {Szyma{\'n}ski}, M.~K. and {Soszy{\'n}ski}, I. and {Ulaczyk}, K. and {Pawlak}, M. and {OGLE Collaboration} and {Abe}, F. and {Asakura}, Y. and {Barry}, R. and {Bennett}, D.~P. and {Bhattacharya}, A. and {Donachie}, M. and {Evans}, P. and {Fukui}, A. and {Hirao}, Y. and {Itow}, Y. and {Koshimoto}, N. and {Li}, M.~C.~A. and {Ling}, C.~H. and {Masuda}, K. and {Matsubara}, Y. and {Muraki}, Y. and {Nagakane}, M. and {Ohnishi}, K. and {Ranc}, C. and {Rattenbury}, N.~J. and {Saito}, To. and {Sharan}, A. and {Sullivan}, D.~J. and {Sumi}, T. and {Suzuki}, D. and {Tristram}, P.~J. and {Yamada}, T. and {Yamada}, T. and {Yonehara}, A. and {MOA Collaboration}},
        title = "{OGLE-2016-BLG-0263Lb: Microlensing Detection of a Very Low-mass Binary Companion through a Repeating Event Channel}",
      journal = {\aj},
         year = 2017,
        month = oct,
       volume = {154},
       number = {4},
          eid = {133},
        pages = {133},
          doi = {10.3847/1538-3881/aa859a},
archivePrefix = {arXiv},
       eprint = {1708.02727},
 primaryClass = {astro-ph.EP},
       adsurl = {https://ui.adsabs.harvard.edu/abs/2017AJ....154..133H}
}

@software{jax18,
  author = {James Bradbury and Roy Frostig and Peter Hawkins and Matthew James Johnson and Yash Katariya and Chris Leary and Dougal Maclaurin and George Necula and Adam Paszke and Jake Vander{P}las and Skye Wanderman-{M}ilne and Qiao Zhang},
  title = {{JAX}: composable transformations of {P}ython+{N}um{P}y programs},
  url = {http://github.com/jax-ml/jax},
  version = {0.3.13},
  year = {2018},
}

\end{document}